\documentclass[aps,prl,superscriptaddress,twocolumn,floatfix]{revtex4-2}
\usepackage{units}
\usepackage{amsmath}
\usepackage{amssymb}
\usepackage{graphicx}
\usepackage{bm}
\usepackage{multirow,xcolor,relsize,ulem,microtype}
\usepackage{braket}

\newcommand{\be}{\begin{equation}}
\newcommand{\ee}{\end{equation}}

\newcommand{\re}[1]{\text{Re}[#1]}

\begin{document}

\title{Twisted light from topological chiral exceptional points in a nanolaser array}

\author{Kaiwen Ji}
\affiliation{Laboratoire Photonique Num\'erique et Nanosciences, Institut d'Optique d'Aquitaine, Universit\'e Bordeaux, CNRS, Rue Fran\c{c}ois Mitterrand, Talence, 33400, France}

\author{Melissa Hedir}
\affiliation{Laboratoire des technologies de la microelectronique, CNRS, Grenoble, 38054, France}

\author{Qi Zhong}
\affiliation{Department of Electrical and Computer Engineering, Saint Louis University,  Saint Louis, MO 63103, USA}

\author{Ramy El-Ganainy}
\affiliation{Department of Electrical and Computer Engineering, Saint Louis University,  Saint Louis, MO 63103, USA}

\author{Alejandro M. Yacomotti}
\email[]{alejandro.giacomotti@institutoptique.fr}
\affiliation{Laboratoire Photonique Num\'erique et Nanosciences, Institut d'Optique d'Aquitaine, Universit\'e Bordeaux, CNRS, Rue Fran\c{c}ois Mitterrand, Talence, 33400, France}

\date{\today}

\author{Li Ge}
\email{li.ge@csi.cuny.edu}
\affiliation{\textls[-18]{Department of Engineering Science and Physics, College of Staten Island, CUNY, Staten Island, NY 10314, USA}}
\affiliation{The Graduate Center, CUNY, New York, NY 10016, USA}

\begin{abstract}
We propose and experimentally demonstrate an orbital angular momentum (OAM) nanolaser array arranged in a ring geometry on an InP-based photonic crystal membrane. The device realizes a non-Hermitian extension of the Rice-Mele model, featuring alternating coupling strengths and imaginary on-site detunings. This configuration supports a symmetry-protected zero mode stabilized by non-Hermitian particle-hole symmetry, which enforces a uniform $\pi/2$ phase shift between adjacent nanolasers—establishing a coherent phase winding around the array. By adjusting the gain/loss contrast in a parity-time (PT)-like pumping scheme, the system can be tuned to a chiral exceptional point, where energy flows unidirectionally between nanocavities despite their reciprocal coupling. This symmetry-enforced, directional tunneling leads to far-field emission carrying non-zero OAM, providing a direct signature of the phase-structured lasing mode. Our results demonstrate a robust and scalable strategy for engineering compact, phase-locked laser arrays with controllable angular momentum output, and open new avenues for structured light generation in integrated photonic platforms.
\end{abstract}

\maketitle

\section{Introduction}

Controlling the angular momentum of light at the micro- and nano-scale has become a subject of intense investigation in recent years. Standard methods for generating non-zero orbital angular momentum $l$ (OAM) typically involve passive phase modulation of an external light source using spiral phase plates, metasurfaces, or spatial light modulators (SLMs) \cite{Karimi2014LSA,Jin2016SR,Sroor2020NPho,chen_highly_2024}. A more integrated approach utilizes compact active devices (e.g.,  lasers) that do not rely on external phase control elements. Examples in this approach include photonic crystal (PhC) lasers based on the bound state in the continuum \cite{huang_ultrafast_2020} and microring lasers based on whispering gallery modes \cite{forbes_orbital_2024}. In the former, however, OAM is not generated deterministically. In the latter, light circulates in clockwise (CW) or counter-clockwise (CCW) directions, and due to the continuous rotational symmetry of such devices, modes with opposite handedness but equal $|l|$ are degenerate and tend to be excited simultaneously or randomly, leading to the cancellation or fluctuation of the net OAM. 

A similar issue arises in arrays of coupled microcavities laser, such as micropillars arranged along a ring
\cite{Sala2015PRX}.
In such systems with discrete rather than continuous rotational symmetry, OAM is no longer a good quantum number, yet states featuring a dominant angular momentum $|l|$ still exist.
Similar to the microring case though, here degenerate states with opposite handedness come in pair, and as a result, selective excitation of only one chiral mode necessarily requires highly specialized pumping schemes. For example, circularly polarized optical pumping has been employed \cite{Zambon2019NPho,Zambon2019OL}, which poses significant challenges for the realization of electrically pumped implementations.

Surprisingly, significant progress towards addressing these challenges came from an unexpected research direction, namely that of non-Hermitian photonics \cite{Feng2017NPho,Ganainy2018NP,Ozdemir2019NM,Miri2019S,Ganainy2019CP,qi_paritytime_2019}. One of the hallmarks of non-Hermitian Hamiltonians is the presence of spectral singularities called exceptional points (EPs) \cite{Heiss2004CJP,Muller2008JPA}. In particular, chiral EPs (CEPs) in microring resonators support a single eigenmode with unambiguous handedness, circulating in one direction without a degenerate counter‑rotating partner. This is achieved by asymmetric scatterings between two counter‑rotating waves of the same $|l|$, which can be realized in microring and microdisk lasers by complex refractive-index modulations along the ring perimeter \cite{Feng_chiral,Miao2016S}, an S-bend \cite{Hayenga2019ACSP}, or judiciously placed nanoscatterers \cite{Yang_chiral,Wiersig2014PRL}. The same principle was realized in a PhC disk with spatially selective pumping \cite{chen2025observation}. More recently, the concept of exceptional surfaces \cite{Zhong2019PRL} has been exploited to realize on‑chip chiral lasers \cite{Liao2023SA}, and a spin–orbit microlaser—comprising two microcavities coupled via a non‑Hermitian synthetic gauge field—has demonstrated emission of light with six degrees of freedom \cite{Zhang_Nature}, firmly placing vortex microlasers at the forefront of photonics research.

Despite these advancements,
microring devices remain fundamentally limited in terms of miniaturization. At telecommunication wavelengths, typical microring resonators span lateral dimensions on the order of tens of micrometers. As such, realizing compact light sources capable of emitting OAM beams remains a significant challenge, particularly at the nanoscale. To address this issue, alternative  platforms are being explored. One promising approach involves the use of topological disclination cavities, which enable controlled OAM emission through rotational symmetry defects engineered at the photonic level \cite{Hwang2024NPho}. 

Another compelling avenue leverages coupled PhC nanocavities, a more mature and widely adopted platform that offers high design flexibility and strong light confinement—making them strong candidates for on-chip OAM laser sources. Recent demonstrations of electrically driven parity-time (PT) symmetric lasers in PhC nanocavities \cite{Takata:21} highlight the significant technological promise of nanolasers with engineered in-situ gain and loss profiles. Moreover, EPs occurring above the lasing threshold have been successfully realized in PT-symmetric PhC cavities \cite{ji2023tracking}. Yet, to date, there is no guiding principle to achieve an OAM laser using a small PhC nanocavity array, and the notion of CEP in PhC geometries remains elusive: Light tunnels between localized modes in neighboring PhC cavities, which is in stark contrast to the traveling waves in a microring resonator. As a result, a ``system-wise'' approach aiming at tailoring the asymmetric scattering of light in the whole system is ineffectual (see Supplementral Note 1).

In this work, we present and experimentally demonstrate an OAM nanolaser array arranged along a minute ring on an InP-based PhC membrane, where lasing occurs in a symmetry-warranted non-Hermitian zero mode  \cite{Schomerus2013OL,St-Jean2017NPho,Zhao2018NC,Parto2018PRL,pan_photonic_2018}. In this zero mode, light tunnels unidirectionally between the nanocavities despite their reciprocal couplings, and the non-zero OAM carried by light is observed in far-field emissions. The guiding principle that enables the chiral modes in our system is the consecutive $\pi/2$ phase jump along the array, which arises from the interplay between non-Hermitian particle-hole (NHPH) symmetry \cite{zeromodeLaser,defectState,Gong} and broken geometric chirality. The light amplitude in each nanolaser also becomes the same when this non-Hermitian zero mode evolves to a CEP.

This guiding principle points to an ``element-wise'' approach, where we construct our array by first identifying the ideal unit cell that repeats periodically.
The result features a modulated gain/loss distribution along this nanolaser array as well as the coupling coefficients, which can be viewed as a non-Hermitian Rice-Mele model 
\cite{Rice1982PRL} with complex detunings.
This guiding principle makes our approach broadly adaptable to other platforms, such as plasmonic lasers, mechanical oscillators, or any system composed of coupled resonant elements. And the periodic nature of them also provides a glimpse into their resilience to system parameters from the band topology perspective, which we characterize using an $\mathbb{Z}_2$ index with non-Hermitian chiral symmetry \cite{Schomerus2013OL,NHChiral}. 

\section{Results}

\begin{figure}[t]
    \includegraphics[width=\linewidth]{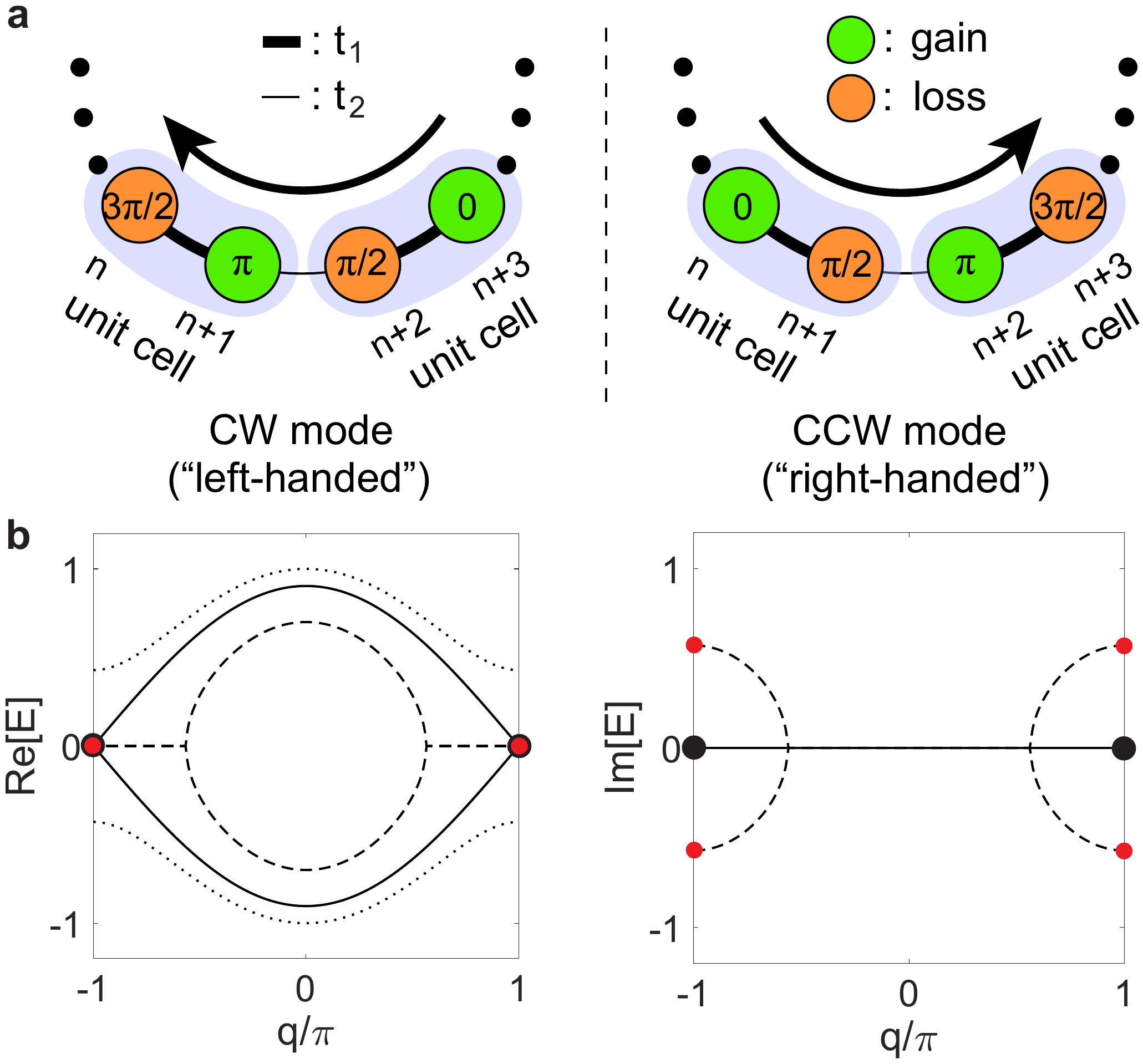} 
    \caption{(a) Schematic diagram of our nanolaser array. The unit cell is a dimer with sufficient gain (loss) in the left cavity, and with the lattice momentum $q=\pi$ between two unit cells, a CCW (CW) non-Hermitian zero mode is achieved (indicated by the arrow and the phase of the complex amplitude in each cavity). Dashed line in the middle shows the mirror symmetry between these two configurations.     
    (b) Band structure of the lattices in (a). Three cases are shown: $\Delta = 0$ (dotted line),  $\Delta=it_1$ (solid line), $\Delta=i(t_1-t_2)$ (dashed line). At $q=\pi$, they have 0, 2 (red dots), and 1 non-Hermitian zero mode (black dot; a CEP). Red and black dots overlap in the left panel. $t_1=5$, $t_2=2$ are used.}
    \label{Fig-Model}
\end{figure}

\noindent
\textbf{Theoretical concept---}The nonzero OAM in our system is enabled by the consecutive $\pi/2$ phase change along the PhC nanolaser array. This guiding principle is largely model independent, and we only require a tight-binding description of our nanolaser array, where each nanolaser is represented by a lattice site and coupled to its nearest neighbors. 

Denoting $\psi_n,\theta_n$ as the complex light amplitude and its phase at the $n$th site, the flux from cavity $n$ to its neighbor cavity $n+1$ is given by 
\cite{ge_optical_2017,ge_non-hermitian_2023} (for completeness, we derive the flux formula in SM note 2):  
\be
J_{n+1,n} =2t_{n,n+1}|\psi_{n+1}\psi_n|\sin(\theta_{n+1}-\theta_n),
\ee
where $t_{n+1,n}=t_{n,n+1}>0$ are the reciprocal couplings between these two cavities. Note that when the field amplitudes are fixed, the magnitude of the flux is maximized when 
\be 
\delta\theta = \theta_{n+1}-\theta_n = \mp\pi/2,\label{eq:DeltaTheta}
\ee
where the sign determines the \textit{local} chirality of wave propagation, i.e., whether the flux is in the CW or CCW direction (see Fig.~\ref{Fig-Model}a).

Unlike passive devices (such metasurfaces) where such a phase change can be engineered using different optical path lengths \cite{chen_highly_2024}, realizing a particular relative phase in a nanolaser array is much more challenging and requires a completely different mechanism. To fulfill the phase requirement (\ref{eq:DeltaTheta}), we first utilize a special property of a non-Hermitian zero mode $\psi$ warranted by NHPH symmetry. This non-Hermitian zero mode \textit{does not} need to be an EP and it is defined by $\re{E}=0$, set at the lasing frequency of a single nanolaser. Denoting it by $\psi=[\psi_1,\psi_2,\ldots]^T$, NHPH symmetry dictates that \cite{zeromodeLaser} 
\be 
C\mathcal{K}\psi=\psi, 
\ee
where $C=\text{diag}[1,-1,1,-1,\ldots]$ is the sublattice operator and $\mathcal{K}$ is the complex conjugation. As a result, a $\psi$ without a node alternates between real and imaginary values, meaning that the phase difference between two neighboring sites is $\pm\pi/2$. NHPH, which is essential here, requires a non-uniform distribution of imaginary detuning or potential in the array. While this restriction is difficult to maintain in a semiconductor laser array, we aim to achieve it at our operating point. Thanks to the robustness of our approach, chiral modes still exist near this point, e.g., when the detunings become complex.

It can be shown that the dimensionless OAM number is directly related to the optical flux with the normalization $\sum_n|\psi_n|^2=1$ (see Supplementary Note 2):
\be
\langle L_z\rangle 
 = \frac{1}{2}\sum_n \frac{J_{n+1,n}}{t_{n+1,n}}
 = \sum_n |\psi_{n+1}\psi_n|\sin\delta\theta.
\label{eq:L}
\ee
To achieve a chiral state and a nonzero OAM, we enforce the same sign of $\delta\theta$ introduced in Eq.~(\ref{eq:DeltaTheta}) on the entire array. For this purpose, we first choose the simplest unit cell satisfying NHPH symmetry, e.g., a dimer with gain and loss. With sufficient gain (loss) in the left cavity, we find $\delta\theta=\pi/2$ ($-\pi/2$) in a non-Hermitian zero mode (see Fig.~\ref{Fig-Model}a). 
By repeating this pattern in an array and choosing a relative phase of $\pi$ between two neighboring dimers (i.e., with a lattice momentum of $q=\pi$), a consecutive phase change of $\pi/2$ ($-\pi/2$) is achieved in the entire array. As a result, $|\langle L_z\rangle|$ reduces to the summation  $\sum_n|\psi_{n+1}\psi_{n}|$, which is maximized when the light amplitude $|\psi|$ is a constant along the ring, achieved at an EP of the dimer thanks to its PT symmetry \cite{bender1998real,Bender1999JMP}.  

\begin{figure}[t]
    \includegraphics[width=\linewidth]{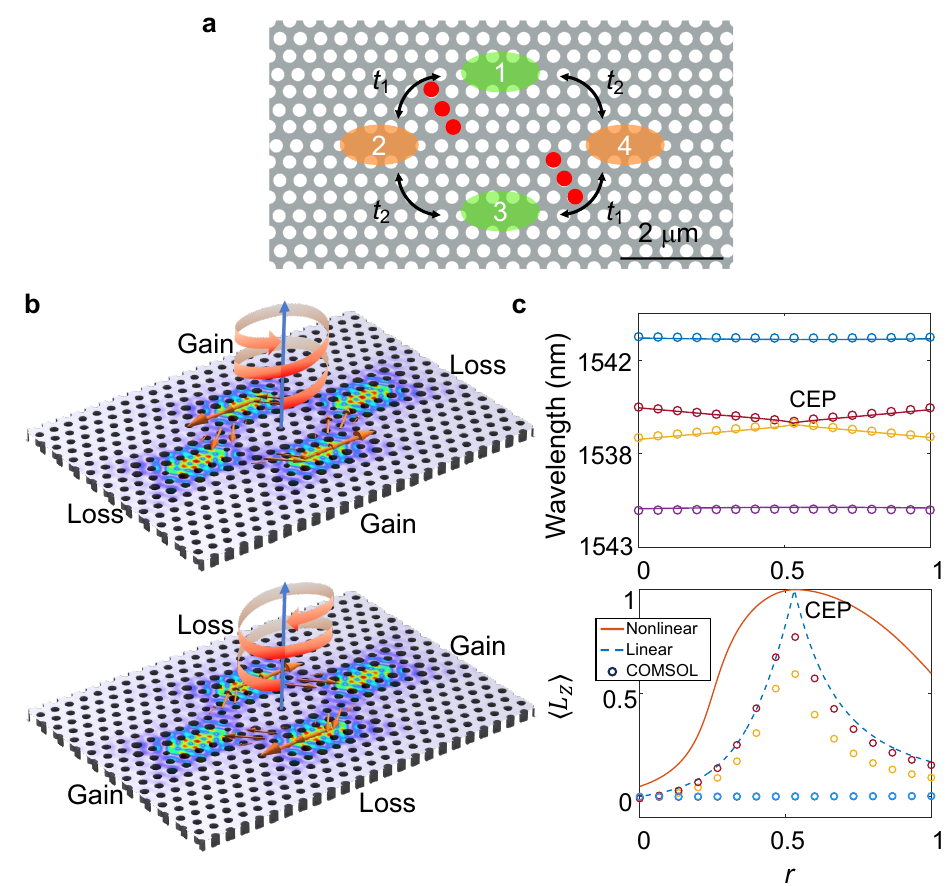}
    \caption{(a) PhC nanolaser array design consisting of four coupled cavities in a ring configuration. Arrows indicate the couplings $t_1\neq t_2$ due to the hole radius change (red dots), and false colors indicate the two sublattices A (cavities 1,3) and B (cavities 2,4) with detuning $\pm\Delta$. (b) Schematic of two chiral modes operating at CEPs in our coupled nanolaser array, with arrows indicating the Poynting vector. (c) COMSOL numerical simulations (circles), eigenvalue calculations with unsaturated gain [dashed blue line, from Eq. (\ref{eq:L})] and nonlinear results including gain saturation (solid red line) showing a CW CEP at $r=0.53$. Here $t_1-t_2=1.2, \alpha=3, \Delta_0=\alpha(t_1-t_2)$ and the total pump power for the linear case is $P_\text{tot}=2.9P_0$, for the nonlinear calculation, $P_\text{tot}=4.9P_0$, where $P_0$ is the threshold for a single cavity.} \label{fig:OAM}
\end{figure}

This approach has two requirements. First, the number of nanocavities needs to be an integer time of four to satisfy the periodic boundary condition in a finite ring, and the most compact one is a ``quartet'' (see Fig.~\ref{fig:OAM}) which we use for our experimental demonstration. Second, we require distinct intra-cell coupling $t_1$ and inter-cell coupling $t_2$. This difference between $t_1$ and $t_2$, together with the modulated gain/loss in NHPH symmetry, breaks the geometric chirality of our array and, in turn, the degeneracy of CW and CCW modes.\\

\noindent
\textbf{Models---}
Having explained the guiding principle that leads to our OAM nanolaser array, we now present the exact model and its rigorous analysis. 
Below we start our analysis by considering the Bloch Hamiltonian of an infinite array, which describes our fintie array at discrete values of the lattice momentum $q$. We use $\Delta(=i|\Delta|),-\Delta$ to denote gain and loss, and we focus on the CCW configuration in Fig.~\ref{Fig-Model}a. Its chiral partner, i.e., the CW configuration in Fig.~\ref{Fig-Model}a, is realized by a complex conjugate that exchanges gain and loss. 

The Bloch Hamiltonian is given by the tight-binding model:
\begin{equation}
    h(q)=
    \begin{pmatrix}
        \Delta   & -t_1-t_2 e^{iq}  \\
        -t_1-t_2 e^{-iq}     &  -\Delta
    \end{pmatrix}. 
\end{equation}
Here $q\in[-\pi,\pi]$ is the lattice momentum and 
we choose $t_1>t_2$ without loss of generality. 
We also note that $q=-\pi$ and $\pi$ are equivalent due to the periodicity of the Brillouin zone, and we have set the lattice constant to be 1. The two complex energy bands of $h(q)$ are given by: 
\begin{equation}
    E_\pm(q)=\pm \sqrt{(t_1^2+t_2^2+2 t_1 t_2 \text{cos}\,q)+\Delta^2}. \label{eq:spectrum}
\end{equation}
Two non-Hermitian zero modes exist at $q=\pi$ when $|\Delta|> (t_1-t_2)$, i.e.,  
$E_\pm(\pi)=\pm i\sqrt{|\Delta|^2 - (t_1-t_2)^2}$ (red dots in Fig.~\ref{Fig-Model}b),
with the wave functions:  
\be 
|\psi_\pm\rangle \propto [1,\,e^{i\xi_\pm}]^T,\quad\xi_\pm\equiv-i\ln\frac{\Delta-E_\pm(\pi)}{t_1-t_2}. \label{eq:psi_pm}
\ee 
Because their $|E_\pm(\pi)|<|\Delta|$, the argument of the log function is a positive imaginary number, leading to $\re{\xi_\pm} = \pi/2$. Therefore, indeed these two modes both have a CCW chirality as depicted by the right panel in Fig.~\ref{Fig-Model}a. They coalesce into a CEP at $\Delta=i(t_1-t_2)$, with which $E_\pm=0$ (black dots in Fig.~\ref{Fig-Model}b) and $\xi_\pm=\pi/2$, leading to $|\psi_\text{EP}\rangle \propto [1,\,i]^T$ and a uniform intensity across the nanolaser array. The CEP in the opposite CW direction is achieved by switching gain and loss (Fig.~\ref{Fig-Model}a), i.e., at $\Delta=-i(t_1-t_2)$.

For our finite array with four cavities, the four modes of our quartet are also described by Eqs.~(\ref{eq:spectrum}) and (\ref{eq:psi_pm}), with the two chiral ones still given by $q=-\pi$ and the other two by $q=0$. The wave functions of the former, now with two unit cells (i.e., $|\psi_\pm\rangle \propto [1,\,e^{i\xi_\pm},-1,\,-e^{i\xi_\pm}]^T$), feature an OAM given by (Supplementary Note 2)
\be 
\langle L_z\rangle_\pm 
=\frac{\sin(\text{Re}\,\xi_\pm)}{\cosh(\text{Im}\,\xi_\pm)}, \label{eq:Lz}
\ee 
which are the same by noticing $\xi_+ + \xi_-=\pi$. 
The other two modes are non-chiral simply as a consequence of $q=0$ (Supplementary Note 2).
 We note that these results hold even when
we allow $\Delta$ to be complex (see the details in the next section), which replaces the system's NHPH symmetry with a non-Hermitian chiral symmetry (Supplementary Note 3). In the next section, we will show how to restore an imaginary $\Delta$ at an CEP, where the OAM is maximized. \\

\noindent 
\textbf{Implementation and Experiments---}
Experiments are implemented on a III-V semiconductor photonic crystal (PhC) platform. PhCs enable extreme device miniaturization and, at the same time, a large control over the system parameters. Here we realize a Rice-Mele array with tailored couplings by employing photonic barrier engineering \cite{haddadi2014photonic}; furthermore, quantum well (QW) InGaAs/InGaAsP materials provide optical gain that can be locally induced by means of spatial light modulation of the pump beam, eventually leading to the imaginary detunings. We fabricated four coupled PhC nanolasers in an InP suspended membrane (Fig.~\ref{fig:OAM}a) and used PhC L3 nanocavities as the building blocks \cite{hamel2015spontaneous}. The leading evanescent coupling takes place along the $\pm60^\circ$ direction, whereas horizontal and vertical coupling can be neglected, therefore forming a four-coupled nanocavity ring. The different coupling coefficients $t_1$ and $t_2$ are designed using a photonic barrier engineering technique, which consists in modifying the size of the nanoholes between the cavities (see the red holes in Fig.~\ref{fig:OAM}a). The measured coupling difference in our device is $t_1-t_2\approx0.07\,\mathrm{THz}$.

The underlying triangular lattice-geometry features a real sublattice detuning $\Delta_0$ between the resonant frequencies of cavities 1,3 and 2,4, which we denote by sublattice A and B. In our PhC geometry, the detuning between two sublattices is measured as $\omega_B-\omega_A\approx-0.5\,\mathrm{THz}$. Importantly, we utilize $\Delta_0=(\omega_A-\omega_B)/2$ to cancel pump-induced blue shift of the nanolaser frequency to reach a CEP, where this blue shift is due to the semiconductor quantum wells in our PhC membrane \cite{ji2023tracking}. 
More specifically, the gain/loss induced complex frequency shift on sublattice $j$ can be written as $(i+\alpha)g_j$, where  
$\alpha$ is the Henry factor and $g_j$ is the gain (if $g_j>0$) or loss (if $g_j<0$) in sublattice $j$. Consequently, $h(q)$ acquires a complex energy shift of $(\omega_A+\omega_B)/2+(i+\alpha)(g_A+g_B)/2$, and $\Delta$ becomes complex in general:
\be 
\Delta = \Delta_0 + (i+\alpha)(g_A-g_B)/2.
\ee
An imaginary $\Delta$ (and NHPH symmetry) is restored at a CEP and $\Delta=\pm i(t_1-t_2)$, achieved with $g_A-g_B = \pm2(t_1-t_2)$ and a passive detuning $\Delta_0=\mp\alpha(t_1-t_2)$. In our particular PhC geometry, we have designed $t_1>t_2$ such that $\Delta_0\approx\alpha(t_1-t_2)>0$, leading to $\Delta=-i(t_1-t_2)$ at the CEP. Therefore we can only efficiently excite the CW CEP as the pump intensity is larger in sublattice B (see Fig. \ref{fig:OAM}b, bottom).

Guided by the non Hermitian chiral zero-mode concept, we first performed a finite-element simulation (COMSOL Multiphysics\textsuperscript{\textregistered}) to identify a CW CEP in a sample with a positive $\Delta_0$ (Fig.~\ref{fig:OAM}c). We also calculated the OAM perpendicular to the PhC $\langle L_z\rangle$ using the near field distribution (Fig.~\ref{fig:OAM}c). Although it differs slightly in the central two modes, their minimum and maximum values at, respectively, $r=0$ and $r=0.53$ (CW CEP), follow the analytical result given by Eq. (\ref{eq:Lz}) (Supplementary Note 4). 

\begin{figure}[b]
    \centering
    \includegraphics[width=\linewidth]{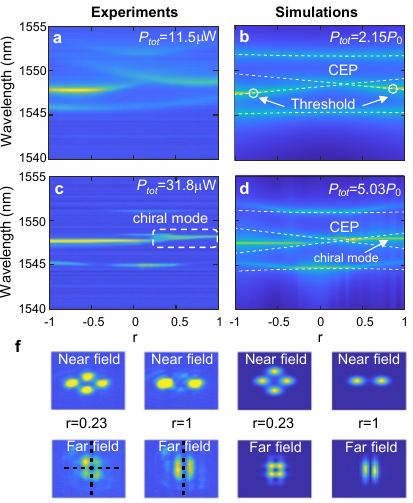}
    \caption{Experimental (left column) and simulation results (right column) of the lasing spectrum. The CEP is below the laser threshold in (a,c) and above in (b,d). Bottom panels f show the field profiles of the chiral mode, with the nodal line(s) added in the far fields of the experimental data. In the nonlinear stochastic simulations, we take $t_1-t_2=0.5$, other parameters can be found in the Supplementary Note 5. The near fields are measured by imaging at the sample plane, whereas the far fields are obtained by imaging at the back focal plane of the objective.}
    \label{Fig-Experiments}
\end{figure}

Here the gain/loss $g_j$ is parametrized by $r=(P_B-P_A)/(P_B+P_A)$, and $P_j$ is the pump power in sublattice $j$ where both cavities are pumped equally. $g_j$ increases linearly with $P_j$ in the linear regime until nonlinear saturation becomes significant. In the experiment, the pump imbalance $r$ is controlled by a pulsed laser source ($\lambda=785$nm, 100ps duration and 10MHz repetition rate), spatially shaped by a liquid-crystal spatial light modulator (SLM) operating in amplitude mode. The modulated pump is then focused down on the sample through a 100$\times$ magnification 0.95 NA microscope objective. The radiation is then collected by the same objective and spectrally resolved with a spectrometer (Supplementary Note 8). 


We first probed the system by pumping it essentially below the laser threshold (see Supplementary Note 9 for the threshold curve). Figure~\ref{Fig-Experiments}a depicts the photoluminescence spectra as a function of $r$. Of the four modes that can be observed, only the one emitting around $\lambda\sim 1547.6$nm lies above the laser threshold for $r<-0.5$, while all the other modes operate below it, including the CEP. Another mode at a slightly longer wavelength progressively takes over as $r$ is increased towards $r=1$, at which point the emission wavelength becomes $\lambda=1548.2$ nm. Notably, we observe a mode switching at the EP at $r_{EP}=0.24$, which we attribute to the efficient collection of the amplified spontaneous emission and the nearly perfect compensation of pump-induced resonance blueshift and sublattice detuning (i.e., an almost imaginary $\Delta$). 

This behavior below the lasing threshold is captured accurately by our tight-binding model in the linear regime, and we show the resulting parametrized trajectories of its eigenvalues in Fig.~\ref{Fig-Experiments}b (dashed lines). 
We also numerically integrated the full stochastic differential equations (Supplementary Note 5) and Fourier-transformed them to account for spectral mode intensities (color maps in Fig.~\ref{Fig-Experiments}b), showing very good agreement with the experimental results. 

Next, we increase the total pump power which brings the CEP into the lasing (and hence nonlinear) regime. As Fig.~\ref{Fig-Experiments}c shows, the mode clamped near 1547.8 nm on the left is the lasing mode when we pump the A sublattice preferrably. As $r$ increases toward 0, another mode near 1548.4 nm switches on and becomes the leading one at $r=0.14$ before the EP is reached. 
%
Unlike the case where the CEP is below the threshold, here the branching near the CEP shown in Fig.~\ref{Fig-Experiments}c is strongly affected by a nonlinear instability, giving rise to a limit cycle before the system becomes stabilized at a larger $r$ (see Supplementary Note 5). 
Beyond the CEP at $r\sim0.23$, 
a single mode with barely changed wavelength is observed (boxed region in Fig.~\ref{Fig-Experiments}c), which is verified in the full nonlinear dynamical model (Fig.~\ref{Fig-Experiments}d) by incorporating gain saturation, carrier-induced refractive index effects as well as stochastic noises. This suggests that the chiral mode wavelength proves resilient against pump imbalance for $r>r_{EP}$, which is visible as a nearly flat trace inside the box. This can be attributed to gain saturation above laser threshold, which substantially reduces the additional frequency shift of the laser mode beyond the CEP ---i.e., as $r\rightarrow 1$. Namely, an upper bound of the observed laser-mode redshift can be estimated to be as small as $\sim \Delta_0$ (equivalently, $0.2$ nm in Fig.~\ref{Fig-Experiments}d). Accordingly, the wavelength remains approximately pinned within this range, and its chiral properties persist, as we demonstrate in the following paragraphs. Additionally, we assess the role of nonlinearity in chiral emission by comparing $\braket{L_z}$ in both the linear (without gain saturation) and nonlinear (with gain saturation) regimes, as shown in Fig. \ref{fig:OAM}(c). In the absence of gain saturation, $\braket{L_z}$ decreases sharply as $r$ departs from the EP [see the dashed line in Fig. \ref{fig:OAM}(c)]. In contrast, when the gain saturation nonlinearity is introduced, both $\pi/2$ phase difference and intensity contrast between sublattices are preserved, allowing $\braket{L_z}$ to remain high over a broad range of $r$, as illustrated by the solid line in Fig. \ref{fig:OAM}(c).


\begin{figure}[t]
    \centering
    \includegraphics[width=\linewidth]{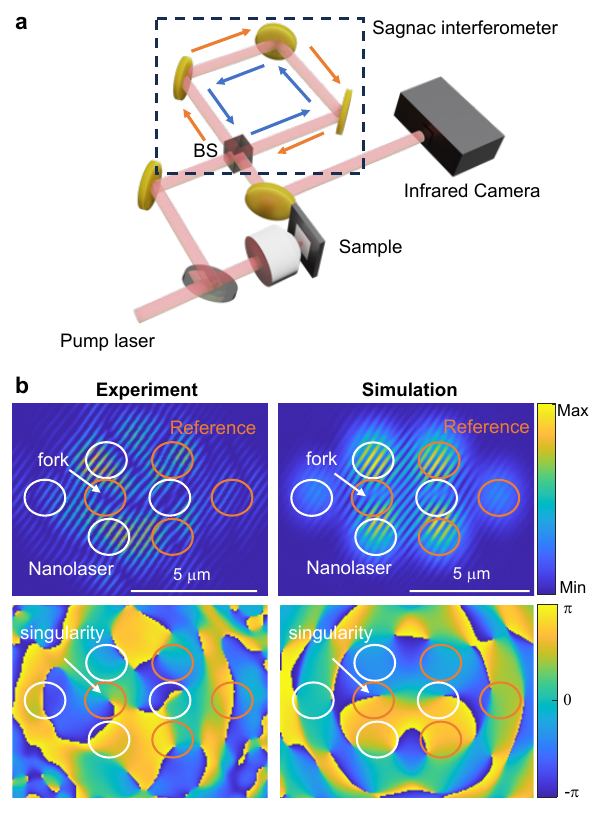}
    \caption{(a) Schematics of the Sagnac interferometer, enabling robust interferometric measurements with spatially shifted patterns. (b) Near-field interference pattern near the EP (left). The circles mark the positions of the cavities, and a fork-shaped feature appears at the center of the structure. The lower panel shows the corresponding phase distribution, revealing a clear phase singularity. The right panel presents the corresponding simulation results.}
    \label{fig:Sagnac}
\end{figure}

In order to investigate the spatial distributions of our chiral mode emission, we first plot the near and far field patterns in Figs.~\ref{Fig-Experiments}c and \ref{Fig-Experiments}d around the CEP and at $r=1$, respectively. As $r$ increases, we observed that the near field evolves from equal intensities in the four cavities to localized distributions in the B sublattice. The relative phase between the cavities is revealed in the far-field pattern: at $r=1$, the far-field shows a single nodal line, from which we know that the wave functions in the two unit cells are $\pi$ out of phase as expected. At the CEP, the far-field shows both a horizontal and a vertical nodal line, indicating that the four cavities are indeed $\pi/2$ out of phase sequentially. 

At the intersection of these two nodal lines lies a nodal point, which is the singularity at the center of our OAM field pattern. 
To visualize this singularity, we implemented a Sagnac interferometer such that two spatially shifted patterns can interfere coherently (see Fig.~\ref{fig:Sagnac}). The spatial phase distribution can be extracted using the off-diagonal Fourier filtering technique, detailed in Supplementary Note 6. The phase map displays a CW winding $2\pi$ at central nodal point of the field, as expected from a CW chiral mode that comes with a positive passive detuning $\Delta_0$.\\


\begin{figure}[h]
    \centering
    \includegraphics[width=\linewidth]{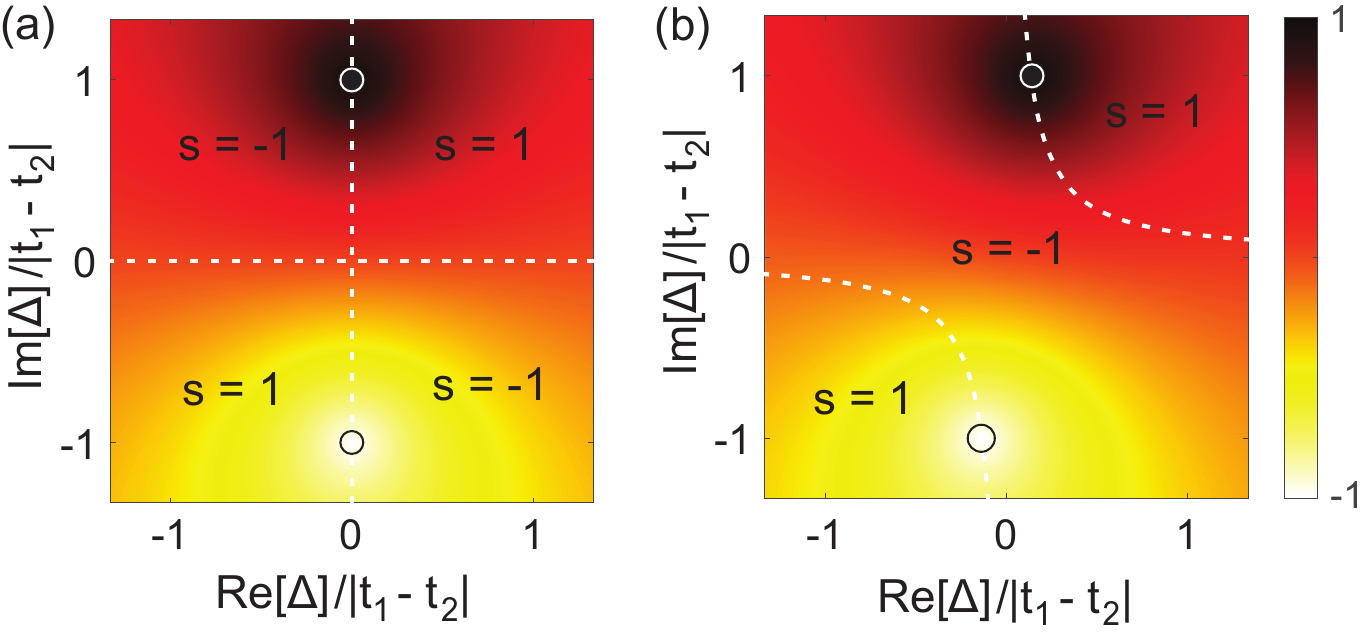}
    \caption{Topological regions characterized by an $\mathbb{Z}_2$ index in the complex $\Delta$ plane. $t_{1},t_2$ are $5,\,2$ in panel (a) and $5-0.32i,\,2+0.1i$ in panel (b). Open circles mark the CEPs, and false colors show the identical OAM $\langle L_z\rangle$ of the two chiral modes in a single quartet [see Eq.~(\ref{eq:Lz})].}
    \label{Fig:topology}
\end{figure}

\noindent
\textbf{Discussions and Conclusion---}
Although systems at an EP are usually considered to be highly sensitive to small perturbations \cite{Wiersig2014PRL}, the chiral state representing the CEP in our system was achieved without meticulous sample preparations. In fact, it is apparent that slight deviations from the exact compensation conditions required by the CEP [$\Delta_0=\alpha(t_1-t_2)$] do not have a strong impact on the OAM of our nanolaser array (Supplementary Note 4).  

Similar robustness involving an EP was discussed from different perspectives, including a bulk-boundary approach \cite{Rivero_robustEP,rivero_robust_2023}, exceptional surfaces \cite{Zhong2019PRL}, and a topological index \cite{Gong}. The latter relies on special symmetry-determined trajectories of energy eigenvalues. In particular, if a positive and a negative eigenvalue of a non-Hermitian Hamiltonian $H$ coalesce at $E=0$ and emerge as two imaginary ones with opposite signs, a $\mathbb{Z}_2$ topological index (e.g., one that takes values of 1 or $-1$) can be constructed using the determinant $s=\text{sgn}\,\text{det}\, (H)$, which changes abruptly from -1 to 1 across the EP at $E=0$.

Because of the complex detuning $\Delta$, however, our energy eigenvalues behave differently. Nevertheless, we can still identify an $\mathbb{Z}_2$ topological index for our CEP:
\be
s = \text{sgn}\,\int_{-\pi}^{0} dq \frac{d}{idq} \text{log}\, \text{det}[U(q)],\label{eq: windingChirality}
\ee
where we have used a polar decomposition \cite{Hall} $h(q)=U(q)P(q)$. $U(q)$ is a unitary matrix and hence its eigenvalues are on the unit circle centered at $E=0$ in the complex plane. So is its determinant, the phase of which is obtained by taking the logarithmic function. Therefore, $\pm1$ are the only two non-zero values of $s$, indicating a CCW (CW) movement of this determinant on the unit circle as $q$ increases from $-\pi$ to 0.

Note that $h(q)$ becomes singular at a CEP where $E=0$. As a result, $U(q)$ is no longer unique and $s$ becomes undefined. In other words, a CEP must appear on the topological boundary that separates two regions with $s=1$ and $-1$. This is shown in Fig.~\ref{Fig:topology}a for a pair of real $t_1$ and $t_2$, where $s$ is calculated as a function of $\Delta$. 
When the couplings $t_1$ and $t_2$ become complex (and hence non-Hermitian) and not in phase, this topological boundary detaches at the origin, but the CEPs still lie on this boundary (Fig.~\ref{Fig:topology}b) \textit{and} feature the strongest OAM, manifesting their resilience on system parameters.

In summary, we presented the first experimental observation of an OAM nanolaser array operating at a CEP, with four PhC cavities arranged on a ring and successive $\pi/2$ phase changes. The resulting phase winding of light in the far-field was evident from the measurement using a Sagnac interferometer. Our element-wise approach to warrant the $\pi/2$ phase changes along the array provides a universal principle for achieving well-controlled OAM in coupled systems, which can be easily extended to states with OAM $|l|>1$ by including $4|l|$ coupled elements (e.g., nanolasers in our PhC membrane). This principle differs drastically from its counterpart in continuous systems, where the system-wise approach aims to induce asymmetric couplings between two counter-propagating modes in ring resonators.  
Although systems operating at EPs are highly sensitive to small perturbations, the OAM emission near the CEP in our system proved to be resilient without meticulous sample preparations. We provided a topological perspective of this robust feature, where the CEP lies on a boundary that separates two distinct regions characterized by an $\mathbb{Z}_2$ index. 

This resilience property in a simple coupling geometry makes our approach broadly adaptable to other platforms, such as plasmonic lasers, mechanical oscillators, or any system composed of coupled resonant elements. Importantly, the small mode volume and high quality factors of PhC cavities make them an ideal platform to study strong interactions between OAM light and matter at an EP, paving the way for building on-chip chiral photonic structures operating in the quantum regime \cite{wang_vortex_2022}.

\textbf{Acknowledgments}
L.G. acknowledges support by the National Science Foundation (NSF) under Grants No. PHY-1847240 and No. DMR-2326698. This work is partially supported by the French National Research Agency (ANR), Grant No ANR-22-CE24-0012-01, by the RENATECH network, and by the ``Grand Programme de Recherche" (GPR) LIGHT.  R.E. acknowledges support from the AFOSR Multidisciplinary University Research Initiative Award on Programmable Systems with Non-Hermitian Quantum Dynamics (Grant No.FA9550-21-1-0202), Army Research Office (W911NF-23-1-0312).

\bibliography{References}

\end{document}